\documentstyle[12pt,epsf]{article}      

\newcommand{\insertplot}[1]{
\begin{center}\leavevmode\epsfysize=8.0cm \epsfbox{#1}\end{center}}

\newcommand{\be}{\begin{equation}}
\newcommand{\ee}{\end{equation}}
\newcommand{\ba}{\begin{eqnarray}}
\newcommand{\ea}{\end{eqnarray}}
\newcommand{\nl}{\nonumber \\}
\newcommand{\vs}{\vspace{0.05cm} }

\begin{document}
\baselineskip 14.5pt
\parindent=1cm
\parskip 3mm


\title{  Hadronization of massive quark matter }

\author{ T.S. Bir\'o, P. L\'evai and J. Zim\'anyi\footnote{Talk given at
the 4th Int. Conference on Strangeness in Quark Matter, July 20-24, 1998,
Padova, Italy}\\[2ex]
KFKI Research Institute for Particle and Nuclear Physics, \\ 
	 P. O. Box 49, Budapest, 1525, Hungary }

\date{31 August 1998}
\maketitle

\begin{abstract}
We present a fast hadronization model for 
the constituent quark plasma ({\bf cqp}) 
produced in relativistic heavy ion collisions at SPS.
The model is based on rate equations and on an 
equation of state inspired by the string phenomenology.
This e.o.s. has a confining character.
We display the time evolution of the relevant physical
quantities during the hadronization process and the
final hadron multiplicities. The results indicate that
the hadronization of {\bf cqp} is fast.
\end{abstract}



\section{INTRODUCTION}
\label{sec1}
 
The production and investigation of quark matter is the main 
motivation of the high energy heavy ion research. In the early days
of the history of this search it was assumed that the 
quark matter formed in high energy heavy ion collisions is similar 
to a quasi-stationary quark gluon plasma, consisting 
of massless gluons and quarks and antiquarks with 
current mass  \cite{Bir1,Raf,Knoll}. However, as more and
more theoretical and experimental results were gathered, 
gradually the picture of a quark matter emerged,
containing effective propagators and interaction 
vertices  \cite{high-TQCD,Kamp,Lev1}. 
In the matter formed in the CERN SPS heavy ion
experiments the effective mass of the interacting gluons 
is larger than that of the dressed quarks  \cite{Lev1}.

On the ground of these theoretical indications we expect
that in the CERN heavy ion experiments not an ideal
 quark gluon plasma,
but a constituent quark antiquark plasma,
 {\bf cqp},
 is formed in some
intermediate state of the reaction \cite{Bir2,Zim3}.

\section{QUARK MATTER WITH STRING-LIKE INTERACTION}

In our model we assume that at the beginning of the hadronization
the matter consists of massive quarks and anti-quarks.
In the time evolution of the system quarks and anti-quarks
form diquarks, anti-diquarks, mesons, baryons and anti-baryons.
We assume that the mixture of all of these particles is in thermal
equilibrium which can be characterized by a temperature.
For the representation of the interaction of the colored particles
we introduce an extra term into the free energy, which is inspired
by the string picture.

As a starting point we consider a mixture of ideal gases of massive
quarks, diquarks, mesons and baryons, and their respective anti-particles.
The corresponding free energy is
\be
F_{{\rm id}} = \sum_i T N_i \left( \ln \frac{N_i}{N_{i,{\rm th}}} - 1 \right)
\, + \, \sum_i m_iN_i,
\ee
with Maxwell-Boltzmann statistics for the non-relativistic
massive matter:
\be
N_i^{{\rm th}}  =  V d_i \int \! \frac{d^3p}{(2\pi)^3} \,
e^{-p^2/2m_iT}.
\ee
Here  $d_i=(2s_i+1)c_i$ are spin and color degeneracy factors.
The chemical potentials in an ideal gas mixture are
\be
\mu_{i,{\rm id}} = T \ln \frac{N_i}{N_{i,{\rm th}}} + m_i.
\ee
The total entropy is given by
\be
S_{{\rm id}} =  \sum_i N_i \left( \frac{5}{2} -  
\ln \frac{N_i}{N_{i,{\rm th}}} \right).
\label{ENTROPY}
\ee
The energy and pressure of such an ideal, non-relativistic mixture
is given by 
\ba
E_{{\rm id}} & = & \sum_i \left( m_i + \frac{3}{2}T \right) N_i, 
\nl
p_{{\rm id}} & = & \sum_i  N_i \frac{T}{V}.
\label{PRESSURE}
\ea

The energy conservation for the expanding system is expressed as

\be
dE + pdV = 0 .
\label{COOL}
\ee

At this point we note an important difference to the canonical
approach to color confinement transition: in an (ideal) mixture of
quarks and hadrons the occupied volume, $V$, is the same for both
components, $V_q=V_h=V$,
the pressure
contributions $p_q$ and $p_h$ are additive. On the other hand
in the application
of the Gibbs criteria of a phase co-existence the volumes $V_q$ and
$V_h$ are additive $V_q+V_h={\rm constant}$, and the partial pressures 
are equal, $p_q=p_h=p$, in phase-equilibrium. 
In our physical picture of hadronization there is
no phase coexistence and the Gibbs criteria do not apply. Colored particles
and color-neutral clusters, pre-hadrons are distributed in a common
reaction volume and chemical reactions convert eventually 
the quark matter into a pure hadronic matter.
\bigskip

The expansion law follows from eq.(\ref{PRESSURE}) and
eq.(\ref{COOL})  as being
\be
\sum_i m_i \dot{N}_i + \frac{3}{2} T \sum_i \dot{N}_i
+ \frac{3}{2} \dot{T} \sum_i N_i + T \frac{\dot{V}}{V} \sum_i N_i = 0.
\label{STAR}
\ee
Due to the foregoing hadronization the number of particles
 decreases, \hbox{($\sum_i  \dot{N}_i < 0$).}
therefore this process re-heats the system. Cooling effects
 are due to
the expansion \hbox{($\dot{V}/V=\partial_{\mu}u^{\mu} > 0$)} and rest mass creation 
\hbox{($\sum_i m_i \dot{N}_i > 0$).
}

In order to ensure the color confinement we supplement
 the model
by the following confinement principle: 
all particles carrying color charge 
(quarks, diquarks, anti-quarks and anti-diquarks) will be penalized
by a free energy contribution stemming from strings.
The number of strings is proportional to 
a weighted sum of the number of color charges,
\be
Q = \sum_i q_i N_i,
\ee
Here  $q_i=0$ for hadrons, 
$q_i = 1/2$ for quarks and anti-quarks, and $q_i = 3/4$ 
for diquarks and anti-diquarks. 
The higher effective charge of diquarks reflects a possibly higher
number of in-medium partners, to which a string is stretched.

The average length $L$ of a string  depends on the
density of colored objects, as $L=n_c^{-1/3}$, where
\be
n_c = \sum_{i=c} N_i/V,
\ee
and the summation $i=c$ excludes color neutral particles (hadrons).
So the free energy of the ideal quark matter - hadron matter mixture,
$F = F_{{\rm id}} + \Delta F,$
is supplemented by the following contribution of strings:
\be
\Delta F = \sigma_s n_c^{-1/3} Q,
\label{DELTA}
\ee
with the effective string tension $\sigma_s \approx 1.0 \ GeV/fm$.

This additional free energy comprises the non-ideality of the equation of
state we use. Since this addition is proportional to the volume $V$ and
the rest depends on densities only, it satisfies thermodynamical
consistency requirements \cite{Toneev} due to its construction.

While there is no new contribution to the
 entropy, $ S = S_{{\rm id}}$,
the pressure, the energy and the chemical potentials of colored
($q_i \ne 0$) particles receive important modifications:
\ba
p     &=& p_{{\rm id}} - \frac{1}{3} \sigma_s n_c^{-1/3} \frac{Q}{V}, \nl
E     &=& E_{{\rm id}} + \sigma_s n_c^{-1/3} Q, \nl
\mu_i &=& \mu_{i,{\rm id}} + \sigma_s n_c^{-1/3} 
(q_i - \frac{1}{3} \overline{q}),
\label{NONID}
\ea
with $\overline{q} = Q/(Vn_c)$. Hadronic chemical potentials have no
modifications at all.

This non-ideal completion of the equation of state influences both
the expansion and cooling and the changes of particle composition.
The correction to eq.(\ref{COOL}) is due to the non-ideal chemical
potentials for colored particles. 
The first principle of thermodynamics for this system 
has the form

\be
TdS = TdS_{{\rm id}} = \left( dE + pdV - \sum_i \mu_i dN_i \right)_{{\rm id}},
\ee
the non-ideal cooling law becomes
\be
 \left( dE + pdV \right)_{{\rm id}} +
\sum_i \left( \mu_i - \mu_{i, {\rm id}} \right) dN_i = 0 \ .
\label{NONCOOL}
\ee
Accordingly eq.(\ref{STAR}) is supplemented by a new generic term
due to the non-ideal equation of state,
\be
\frac{\dot{T}}{T} = - \frac{2}{3} \frac{\dot{V}}{V} 
- \frac{\sum_i \dot{N}_i}{\sum_i N_i} 
- \frac{2}{3}\frac{\sum_i (m_i/T)  \dot{N}_i}{\sum_i N_i} 
-\frac{2}{3}\frac{\sum_i (\mu_i/T - \mu_{i,{\rm id}}/T)\dot{N}_i}{\sum_i N_i}
\ee
The additional term,
\be
\sum_i \frac{\Delta\mu_i}{T} \dot{N}_i =
\sigma_s \sum_{i=c} n_c^{-1/3}(q_i - \frac{1}{3}\overline{q}) \dot{N}_i,
\ee
is negative if color charges become eliminated from the mixture.
Therefore color confinement, causing an extra suppression of
equilibrium numbers of quarks and alike particles, re-heats the
expanding fireball as well as the ``normal'' chemistry of the
ideal quark - hadron mixture.

The only physical effect besides a fast expansion - which has, however,
kinematical limits stemming from scaling relativistic expansion -
that can cool the mixture sufficiently is rest-mass production.

Since in our equation of state we have an explicit interaction energy
between the  quarks, our effective quark masses should be less
than that given in Ref.\cite{Lev1}. We shall use the following values:
$ ( m_u=[m_{u0}^2 + m_{th}^2]^{1/2} $,
$  m_d=[m_{d0}^2 + m_{th}^2]^{1/2} $,
$  m_s=[m_{s0}^2 + m_{th}^2]^{1/2} $,
with thermal mass $ m_{th} = 0.15 $ GeV and
$m_{u0}=m_{d0}\approx 0$, $m_{s0}=0.15$ GeV. 
The clusters have a mass according to the average mass
extra to the summed valence quark masses of the two lowest
lying hadron multiplets: the pseudo-scalar and vector meson nonets
and the baryon octet and decuplet, respectively
($\Delta m_{{\rm mes}} = \Delta m_{{\rm diq}} = 0.408$ GeV,
$\Delta m_{{\rm bar}} = 0.805$ GeV).

\subsection{Initial state}
\label{subs223}

The initial energy density --- distributed along the beam direction
between $-\tau_0 \sinh \eta_0$ and  $\tau_0 \sinh \eta_0$ ---  can be related
to the center of mass bombarding energy $\sqrt{s}$ in the experiment,
\be
\varepsilon_0 = \frac{\sqrt{s}}{\pi R_0^2\tau_0 2 {\rm sh} \eta_0}.
\ee
On the other hand the initial invariant volume dual to $d\tau$
at constant $\tau=\tau_0$ is given by
\be
V_0 = \pi R_0^2\tau_0 2\eta_0.
\ee
The initial internal energy (i.e. the energy without the collective
flow of a fluid cell) at $\tau=\tau_0$ is therefore
less than $\sqrt{s}$ for finite $\eta_0$:
\be
E_0 = \varepsilon_0 V_0 = \frac{\eta_0}{ {\rm sh}\eta_0} \sqrt{s}.
\ee
At the CERN SPS experiment $\eta_0 \approx 1.85$ (due to some stopping),
$R_0 \approx 7$ fm, $\tau_0 \approx 0.5$ fm and
we obtain $V_0 \approx 431$ fm$^3$ and $E_0 \approx 2.13$ TeV.
Compared to the total energy of about $\sqrt{s} = 3.4$ TeV
(carried by about $390$ participating nucleons in a central
Pb-Pb collision)
approximately two third of the energy is invested into
rest mass of newly produced particles and thermal motion and
and one third  into the flow.

Comparing this with an alternative expression for the thermal energy
of an ideal massive quark matter,
\be
E_0 = \sum_i N_i(0) (m_i + \frac{3}{2} T),
\ee
one can estimate the initial temperature at the beginning of hadronization.
Using our standard values for the incoming quark numbers \cite{SQM97}
$N_u(0) = 544,$ $N_d(0) = 626$, further, assuming  that 400 $u \overline u$,
400 $ d \overline d $ and,  with $f_s=0.22 $, 
176 $ s \overline s$ quark anti-quark
 pairs are created in one central collision, 
 we get from the above equation $T_0 = 0.18$ GeV.
We use these numbers for the newly produced quark pairs 
in order to arrive at the experimentally measured hadron and 
strange particle numbers.

\subsection{Hadronization processes}

Chemical equilibrium is not supposed initially, rather a definite
over-saturation of quarks  in the reaction volume.
The initially missing color-neutral hadron states - mesons and
baryons - are formed due to quark fusion processes in a non-relativistic
Coulomb potential. The rates for different flavor compositions
differ mainly due to the different reduced masses of quark  anti-quark
or quark diquark pairs. First of all this influences the Bohr radius
in the Coulomb potential \cite{Bir2}. 
Of course, the presence of a medium -
which establishes the necessary momentum balance after the fusion -
also influences the hadronization rate. The cross section for
such a $2 \rightarrow 1$ process in medium is
\be
\sigma =  \left(\frac{\rho}{a}\right)^3
\frac{16M^2\sqrt{\pi}\alpha^2}{\left( \vec{p}^2 + 1/a^2 \right)^2}
\ee
with $a = 1/(\alpha m)$ Bohr radius of the $1s$ state in the
Coulomb potential and $\rho $ is the Debye screening length
\cite{Bir2}.  Here $\vec{p}$ is the relative momentum of the
hadronizing precursors, $m$ is their reduced mass and $M$ is the
total mass. For $\alpha $ a running coupling constant was 
used.

The relative momenta are taken from a random Gaussian
distribution,
\be
d{\cal P}(\vec{p}) \propto e^{-p^2/2mT} d^3p,
\ee
at temperature $T$ and reduced mass $m$.
This method allows us to simulate thermally averaged
hadronization rates,
\be
R = \langle \sigma \frac{|\vec{p}|}{m} \rangle.
\label{RATE}
\ee

\subsection{Dynamical Effects of Confinement}

Besides the "confining correction" in the equation of state 
 we also apply a {\em dynamic confinement}
mechanism in our model: the medium screening length $\rho$
occurring in the hadronization cross section will be related to
strings pulled by color charges trying to leave the reaction zone.
This way the screening length $\rho$ is increased as the color
density decreases: we keep, however, the product $\rho^3 N_c/V$
constant,
\be
\rho(t) = \rho(0) * \left( N_c(0)/N_c(t) \right)^{1/3}.
\ee
Here $N_c$ stands for the number of colored objects.

\subsection{Reaction network}

What remains to specify the model is the system of rate equations describing
the transformation of quark matter into hadronic matter.
We consider $N_F=3$ light quark flavors $u$, $d$ and $s$.

There are $N_F(N_F+1)/2=6$ possible diquark flavors and the same
number of anti-diquark flavors. The number of quark anti-quark 
flavor combinations is $N_F^2=9$ while that of quark or
anti-quark triplet combinations is $N_F(N_F+1)(N_F+2)/6=10$.
In the hadronizing quark matter we deal with altogether
$2*3+2*6+9+2*10=47$ sorts of particles.

Let us generally denote quarks by $Q$, diquarks by $D$,
mesons by $M$ and baryons by $B$.
The possible fusion reactions are:
$
Q + Q   \longrightarrow  D, 
\overline{Q} + \overline{Q}   \longrightarrow  \overline{D}, 
Q + \overline{Q}   \longrightarrow  M, 
Q + D   \longrightarrow  B, 
\overline{Q} + \overline{D}   \longrightarrow  \overline{B}. 
$

Our model is completed by the system of rate equations.
Considering a general reaction of type
$ i + j \longrightarrow k$
we account for the changes
\be
dN_i = dN_j = - Adt,  \qquad  dN_k = +Adt,
\label{COMP}
\ee
cumulatively in each reaction.  Here
\be
A = R_{ij\rightarrow k}  N_i N_j 
\left( 1 - e^{\frac{\mu_k}{T} - \frac{\mu_j}{T} - \frac{\mu_i}{T} } \right)
\label{BACK}
\ee
with a thermally averaged  rate 
$ R_{ij\rightarrow k}$ (cf. eq.(\ref{RATE})).
The changes stemming from different reactions accumulate to a
total change of each particle sort in a time-step $dt$.

\subsection{Chemical equilibrium}

Chemical equilibrium is defined by the requirement that all chemical
rates vanish (cf. eq.\ref{BACK}). It leads to relations like
\be
  \mu_i^{{\rm eq}} + \mu_j^{{\rm eq}} = \mu_k^{{\rm eq}}
\label{CHEMEQ}
\ee
for each reaction channel. The correspondence between equilibrium
chemical potentials and equilibrium number densities is, however,
in the general case not as simple as for a mixture of ideal gases.
From eq.(\ref{NONID}) we obtain an implicit equation for the
equilibrium numbers,
\be
 N_i^{{\rm eq}} = N_{i, {\rm th}} \, 
e^{\frac{\mu_{i}^{{\rm eq}}-b_i(N^{{\rm eq}})-m_i}{T} }
\ee
with
\be
 b_i(N^{{\rm eq}}) = \left. \sigma_s n_c^{-1/3} \left( q_i - \frac{1}{3} \overline{q} \right)
\right|_{{\rm eq}}
\ee
We call the attention to the fact that this equation does not lead to
a chemical equilibrium state below a critical temperature.

\vs

Applying eq.(\ref{NONID}),
the non-equilibrium chemical potentials, and hence the essential factors,
$e^{-\mu_i/T}$, in the chemical rates (eq.(\ref{BACK})) can be expressed
as
\be
 e^{-\mu_i/T} = e^{-\mu_i^{\rm eq}/T} \cdot 
e^{\frac{b_i(N^{{\rm eq}})-b_i(N)}{T}} \cdot
\frac{N_i^{{\rm eq}}}{N_i}.
\ee
In the combinations appearing in the detailed balance factor of the
rate equations using eq.(\ref{CHEMEQ}) we obtain
\be
 1 - e^{\frac{\mu_k-\mu_i-\mu_j}{T}} = 1 - 
\frac{N_i^{{\rm eq}}}{N_i} \frac{N_j^{{\rm eq}}}{N_j} \frac{N_k}{N_k^{{\rm eq}}}
e^{\frac{\Delta\mu_k-\Delta\mu_i-\Delta\mu_j}{T} }, 
\ee
with
\be
\Delta\mu_i = b_i(N) - b_i(N^{{\rm eq}}).
\ee
The corrections $b_i(N)$ in the non-equilibrium chemical potentials
may in general depend on the number densities of several other
components on the mixture.

\vs
At this point we note that the extra $e^{-\Delta\mu_i/T}$ factors
occur for any non-ideal equation of state where the correction to
the free energy density is a nonlinear function of the number
densities.

\vs

\subsection{Hadronic decays}

The set of rate equations describes the time evolution of the
number of all involved quark and antiquark clusters.
 In order to get the final
hadron numbers we integrate these equations until the
number of colored particles becomes negligible. 
At this time the number of clusters are divided between the 
corresponding hadrons according to the spin degeneracies 
of the multiplets.
This way we obtain a number of hadronic resonances
(in the present version the vector meson octet and baryon
decuplet).
Finally hadronic decays are taken into account with
the dominant
branching ratios obtained from Particle Data Table \cite{PDT}.
We assume that secondary hadron-hadron
interactions have a negligible effect on the
finally observed hadronic composition.
The time evolution of the entropy and temperature
is obtained by simultaneous integration of eqs.(\ref{ENTROPY})
and (\ref{STAR}). 


\section{ Numerical results and discussion}

 For the parameters describing the initial state we used those
given in subsection \ref{subs223}, while 
for the parameters determining the dynamics of the 
hadronization we used the
following values: $\rho = 0.2 fm$, $\alpha = 1.4$.

\insertplot{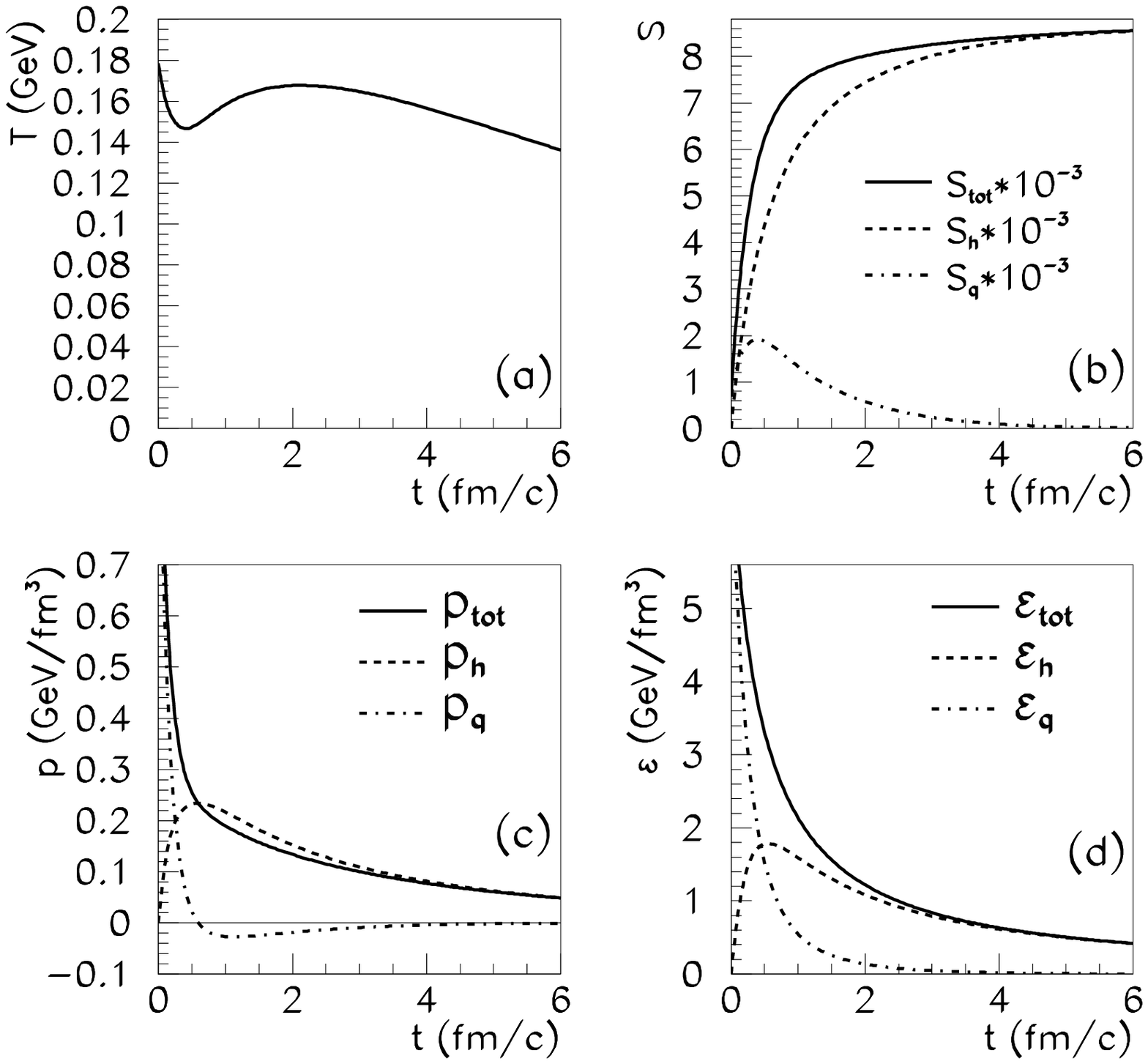}
 \begin{center}
  \begin{minipage}[t]{13.054cm}
  { {\bf Figure 1.}
The time evolution of the temperature $T$ (a), the  
entropy $S$  (b),
the pressure $p$ (c) and the energy density $\varepsilon$ (d)
of the system together with the partial contributions of quarks and hadrons.}
  \end{minipage}
 \end{center}

From Fig.1a we can see, that at the beginning of the hadronization 
there is a rapid decrease in the temperature due to the rest mass formation
of the hadrons. Shortly after that, the re-heating starts as an effect
of color confinement (see eqs.(28) and (32)).
 Fig.1b shows, that the 
total entropy is monotonically increasing during the hadronization.
In Fig.1c one can
observe an interesting pattern in the time evolution of the pressure. 
The partial pressure 
of the interacting cqp rapidly decreases as the number of quarks 
decrease. As the color density drops, this pressure becomes even negative.
The increasing hadron partial pressure, however, over-compensates this negative
value. 
The partial and total internal energy evolution,
displayed in Fig.1d, shows, that the hadronization is completed
at $2$ fm/c after the beginning of the process. The decrease of the 
internal energy is compensated by the 
work of pressure while making the flow. 

Fig.2 shows the time evolution of different colored particles
and color neutral clusters corresponding to
different hadrons. 
The diquarks are produced from the quarks, and then they 
contribute rapidly to the formation of baryons.

\insertplot{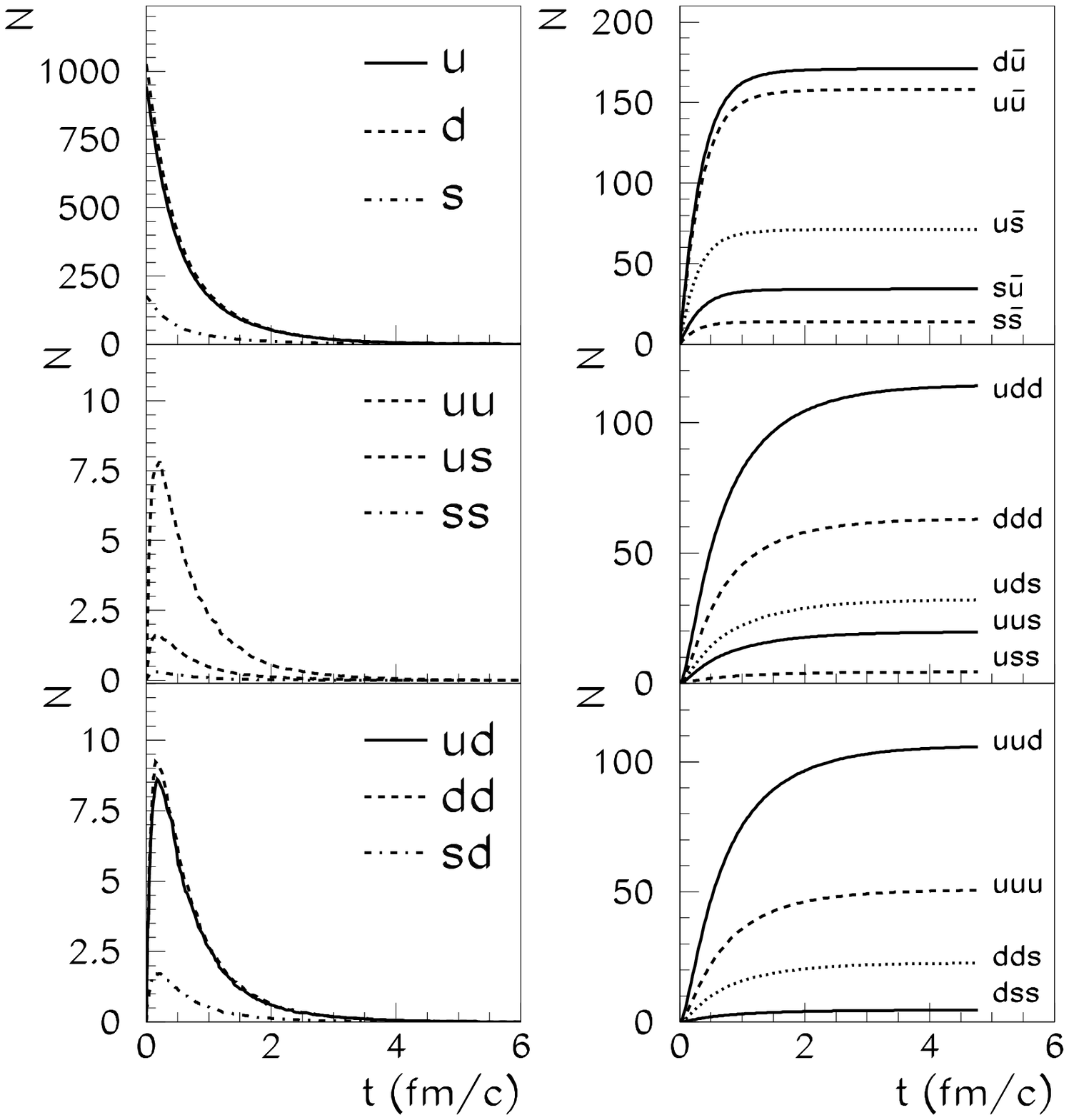}
 \begin{center}
  \begin{minipage}[t]{13.054cm}
  { {\bf Figure 2.}
  The time evolution of the number of colored particles
 and color neutral clusters.
  The line styles of different flavor compositions are
  indicated in the respective figures.  }
  \end{minipage}
 \end{center}

In the ALCOR model the ratio of hadronic species are determined
by the ratio of steepnesses of these curves. Since these curves do not 
cross each other, one can understand, why the algebraic ALCOR 
approach to the solution of rate equations is a good approximation. 

 In Table 1 the hadron numbers obtained with the Transchemistry
model and those obtained with the ALCOR \cite{SQM97}
and the RQMD \cite{RQMD} models are shown 
together with the few published  experimental data.
 Table 2 shows a comparison 
for the multi-strange baryon ratios. While in
many cases there are intriguing agreements, in some other cases 
 there are  some discrepancies.
The reason for this may originate from two sources: i) the
experimental data referred here are the production ratios in
 the overlap window of the detector acceptances.
Thus, if the momentum distribution of the two particle sorts
 is not the same,
then these ratios are not equal to the total number
 ratios. ii) these calculated
values are more sensitive to the simplifying assumption, that 
 the hadronization happens into
the lowest energy baryon octet and decuplet and the two lowest energy
meson octet.

\begin{center}
\begin{tabular}{||c||c||c|c|c||}
\hline {\bf Pb+Pb} & { NA49}
& {TrCHEM.}  & { ALCOR } & { RQMD } \\
\hline
\hline
 $h^{-}$  & $680^a$ 
&677.3    &679.8    & \\
\hline
\hline
 $\pi^+$  &  
&581.5    &590.6    &692.9  \\
\hline
 $\pi^0$  &  
&616.8    &605.9    &724.9  \\
\hline
 $\pi^-$  &  
&613.4    &622.0    &728.8  \\
\hline
 $K^+$    & $76^*$  
&\ 79.5   &\ 78.06  &\ 79.0   \\
\hline
 $K^0$    &  
&\ 79.5    &\ 78.06 &\ 79.0  \\
\hline
 ${\overline K}^0$    &  
&\ 39.6   &\ 34.66  &\ 50.4  \\
\hline
 $K^-$    & $\{32\}^b $  
&\ 39.6   &\ 34.66  &\ 50.4   \\
\hline
\hline
 $p^+$    &  
&159.2    &153.2    &199.7  \\
\hline
 $n^0$    &  
&175.5    &170.5    &217.6  \\
\hline
 $\Sigma^+$    &  
&\ \ 8.4  &\ \ 9.16 &\ 12.9  \\
\hline
 $\Sigma^0$    &  
&\ \ 9.8  &\ \ 9.76 &\ 13.1     \\
\hline
 $\Sigma^-$    &  
&\ \ 9.5  &\ 10.39  &\ 13.3    \\
\hline
 $\Lambda^0$   &  
&\ 46.8   &\ 48.85  & \ 35.3  \\
\hline
 $\Xi^0$  &  
&\ \ 4.35 &\ \ 4.89 & \  \ 4.2   \\
\hline
 $\Xi^-$  &  
&\ \ 4.38 &\ \ 4.93 & \ \ 4.2   \\
\hline
 $\Omega^{-}$  &  
&\ \ 0.42 &\ \ 0.62 &   \\
\hline
\hline
 ${\overline p}^-$   &  
&\ \ 8.96 &\ \ 6.24  &\ 27.9    \\
\hline
 ${\overline n}^0$   &  
&\ \ 9.01 &\ \ 6.24  &\ 27.9   \\
\hline
 ${\overline \Sigma}^-$   &  
&\ \ 1.01 &\ \ 0.91   &\ \ 4.6   \\
\hline
 ${\overline \Sigma}^0$   &  
&\ \ 1.10 &\ \ 0.91   &\ \ 4.6    \\
\hline
 ${\overline \Sigma}^+$   &  
&\ \ 1.00 &\ \ 0.91   &\ \ 4.6    \\
\hline
 ${\overline \Lambda}^0$   & 
&\ \ 5.24 &\ \ 4.59  &\ 10.7  \\
\hline
 ${\overline \Xi}^0$   &  
&\ \ 1.09 &\ \ 1.12  &\ \ 2.0    \\
\hline
 ${\overline \Xi}^+$   &   
&\ \ 1.09 &\ \ 1.12  &\ \ 2.0    \\
\hline
 ${\overline \Omega}^{+}$   &  
&\ \ 0.23 &\ \ 0.35  &   \\
\hline
\hline
 $K^0_{S}$   &$ \{54\}^{b,c}$  
&\ 59.6   &\ 56.36  & \ 63.5  \\
\hline
$p^+-{\overline p}^-$ &$\{145\}^a$ 
&150.2    &147.03   &171.8 \\
\hline
 $\Lambda^0$-like &$\{50\pm 10\}^b$  
&\ 62.30  &\ 69.07  & 56.8    \\
\hline
 ${\overline \Lambda}^0$-like &$\{8\pm 1.5\}^b$  
&\ 8.14   &\ 8.12   & 19.3   \\
\hline
\hline
\end{tabular}
\end{center}
\vskip 0.5truecm
\noindent {\bf Table 1:}
Total hadron multiplicities for $Pb+Pb$ collision
at bombarding energy 158 GeV/nucleon. The displayed
experimental results are from the
NA49 Collaboration:
${}^a$ is from \cite{NA49QM96};
${}^b$ is from \cite{NA49S97}; ${}^c$ is from
\cite{NA49S96}; ${}^*$ is estimated from $\{K^-\}$ and
$\{K^0_S\}$. 
Theoretical results are from
the Transchemistry, ALCOR
and RQMD  
("ropes + no re-scattering" version) model. 
Here it is
$\Lambda^0-$ like $\equiv\Lambda^0+\Sigma^0+\Xi^-+\Xi^0+\Omega^-$.
\bigskip

\begin{center}
\begin{tabular}{||c||c|c|c|c||}
\hline {\bf Pb+Pb} &  {\bf WA97}  & {\bf TrCHEM} & {\bf ALCOR}
 \\
 \hline
 \hline
 ${\overline {\Xi}}^+/\Xi^-$    
 &$0.27 \pm 0.05$ &0.25 & 0.23    \\
 \hline
 ${\overline {\Omega}}^+/\Omega^-$
 &$0.42 \pm 0.12$ &0.55 & 0.56    \\
 \hline
 $\Omega^-/\Xi^-$                
 &$0.19 \pm 0.04$ &0.10& 0.13    \\
 \hline
 ${\overline {\Omega}}^+/{\overline {\Xi}}^+$ 
 &$0.30 \pm 0.09$ &0.21& 0.31   \\
 \hline
 \hline
 \end{tabular}
 \end{center}

\noindent{\bf Table 2.}
Strange baryon and anti-baryon ratios measured by
WA97 Collaboration 
\cite{S97WA97}
and obtained from Transchemistry and ALCOR model
for $Pb+Pb$ collision at 158 GeV/nucl bombarding energy.
The experimental data are the production ratios in the
overlap window of the detector acceptance.
\bigskip


\section{Conclusion}

In this paper we presented a new model for the hadronization
of a quark antiquark plasma ({\bf cqp}) based on rate equations in a
 quark matter - hadron matter mixture.
 The color confinement was taken into account by
using consistently a plausible equation of state 
motivated by the string model.
Our results presented in the Figures clearly show a very fast
hadronization.
Observing the shape of the time evolution of different
hadron multiplicities, it became understandable, why  
the simpler algebraic approximation, applied in the
ALCOR model, works so well. The comparison with the existing
experimental data indicate, that it is possible, that
 in the PbPb collision at SPS a piece of matter is formed, inside which
the massive quarks and anti-quarks interact with a string like
mean field.

Finally we emphasize, that  this type of phenomenological
 investigations
are necessary,  as long as the hadronization of the quark matter 
as a non-equilibrium, non-static,
non-perturbative process, cannot be described  with  other methods.


\section*{Acknowledgments}

Stimulating discussions with J.Knoll, A.A.Shanenko, V.Toneev are acknowledged.
This work was supported by the Hungarian Science Fund grants T024094 and
T019700, and by a common project of the Deutsche Forshungsgemeinschaft
and the Hungarian Academy of Science DFG-MTA 101/1998.

\end{document}